# Metameterial refraction characteristics within the millimeter-wave range

*G.A. Naumenko*[*], *A.P. Potylitzyn, M.V. Shevelev, V.V. Bleko, V.V. Soboleva*

National Research Tomsk Polytechnic University, 634050, Tomsk, Russia

Phase delays, spectral, orientation, and angular dependences of radiation refraction in metamaterial targets have been experimentally studied within the millimeter-wave range. It has been shown that angular and spectral dependences have a complicate structure, which demonstrated properties typical both for positive and negative refraction indexes. At this, a change of refraction nature occurs in radiation wavelength interval comparable with the sizes of metamaterial structural elements.

## 1. Introduction

Recently, metamaterials have attracted a great interest due to their specific physical properties and possibilities of advanced applications [1, 2]. In conventional materials permittivity and magnetic permeability are positive. Nevertheless, for some structures not only permittivity, but also permeability can have negative values. Therefore, phase and group velocities of the electromagnetic wave can be oppositely directed. This results in a number of interesting properties of such materials [3]. These materials are often called double negative or left-handed materials. In 1968 Victor Veselago proved the possibility of existence of physical media with the negative refractive index [4]. Later, this idea was developed by many authors in the frame of macroscopic electrodynamics [5,6,7,8,9,10], where among other things, it was shown a possibility of shunting physical objects' visibility and a possibility of overcoming the diffraction limit.

Negative permittivity and magnetic permeability for frequencies close to resonance frequency can be obtained by creation of such structures as split-ring resonator arrays [11]. Combining these resonators with thin strips the properties of left-handed materials were experimentally shown in [12, 13]. Later, the analytical developments of these material properties were followed, as well as the estimate of numerical modeling with scalar values of refractive index [14, 15]. Experimental measurements were carried out in a waveguide chamber which is limited to one of the types of metamaterial structures [16]. A shortage of this approach is the problem of combining surface wave in a waveguide with wave propagation in the metamaterial structure.

In [17] the interaction between radiation and metamaterials has been discussed in a free space. Such approach has less restrictions for the sizes and orientation of targets in space. However, in the cited work there was considered only radiation transmission through the structures but the problem of its refraction was omitted.

It is worth noting that the problem of the interaction between the radiation and metamaterials may be considered from the similarity criteria which is based on the ratios between the geometric characteristics of the metamaterial structure and the radiation wavelength. Therefore, the results obtained in the terahertz range can be used, for example, in the millimeter-wave range and vice versa. On the other hand, it is easier and cheaper to provide manufacturing of the metamaterial structure for the millimeter-wave range of radiation. This fact has defined the choice of the spectral range by us.

## 2. Experiment

From the experimental results obtained in [11,12,13,18] the phenomenon of negative refractive index is best manifested in the use of the array of loops with slits in them at opposite ends and single vertical strips along them [18]. These structures were chosen for the detailed study of their radiation response functions.

---
[*] E-mail: naumenko@tpu.ru

The investigation of metamaterial properties was started from the study of properties of the radiation transmission through a separate two-dimensional structure (Fig.1).

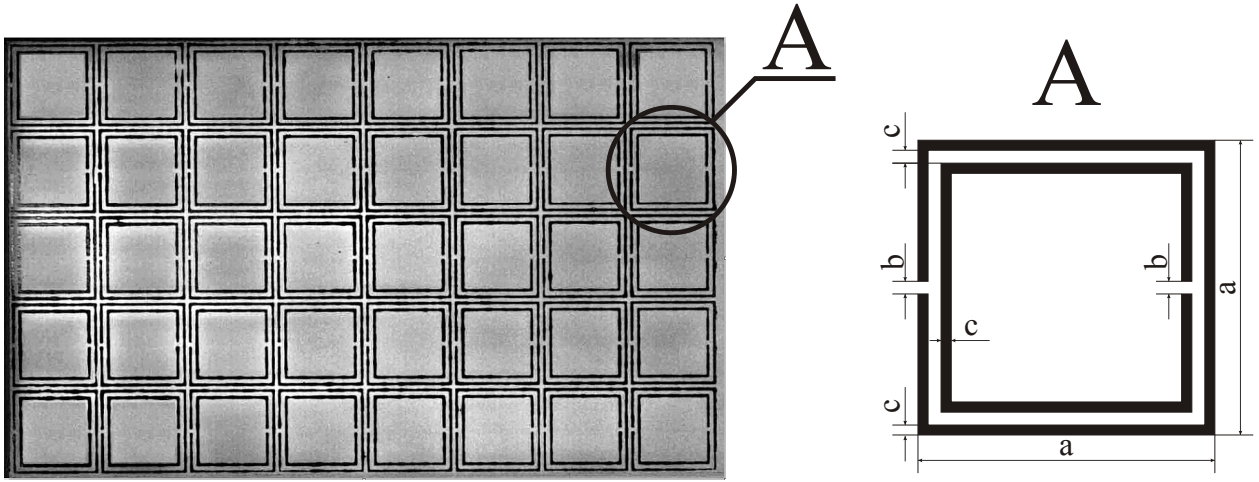

Fig. 1. Configuration of two-dimensional structure of metamaterial

This structure is applied as copper coating on a woven-glass reinforced substrate of 0.5 mm thickness with geometry of *c=b=0.05a* (see insertion in Fig. 1). By analogy with conventional materials where the refractive effect is determined by the ratio between radiation phase velocity in the target and in vacuum, the spectral responses and refraction indexes in the metamaterial target should be detected by a phase delay of radiation at separate plates of the structure. Results obtained for the phase delay on two-dimensional structures of metamaterial with different parameters are given below.

The experimental scheme is shown in Fig. 2. Gunn diode was used as a radiation source which provided two isolated lines in the spectrum ($\lambda_1 = 9.2$ mm, $\lambda_2 = 29$ mm) with controlled intensity.

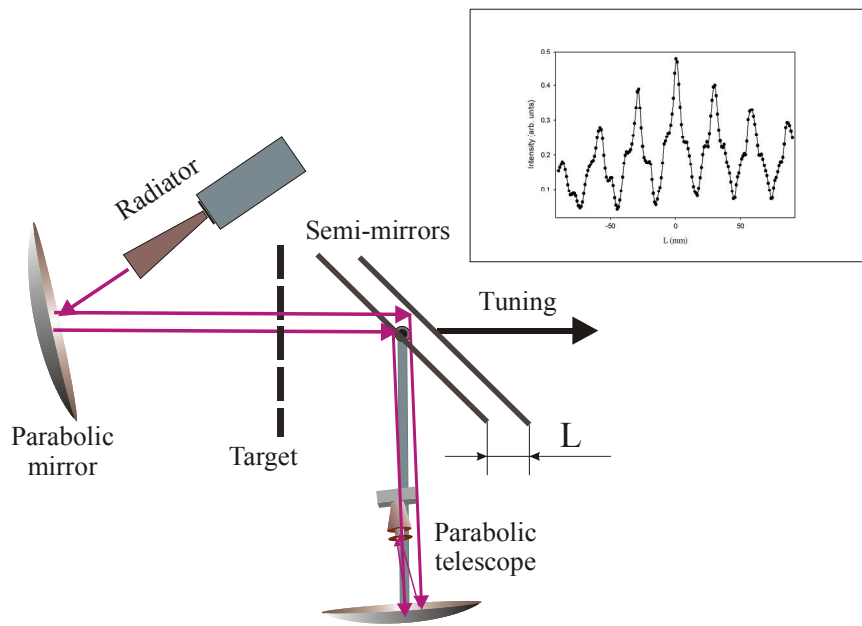

Fig. 2. Scheme of the radiation phase delay measurement with the example of typical interferogram

In the experiment the interferometer was used with wave front division by two mirrors, one of which was movable [19]. Radiation was registered by a detector with characteristics described in [20]. A parallel radiation beam was formed by a parabolic mirror in the focus of which the radiation source was placed. This beam was directed to the interferometer.

For our experimental conditions the effect of prewave zone appears, when not angular but rather spatial radiation distribution is measured and an angular distribution is distorted by the target size. To exclude the effect of prewave zone the radiation was focused on the detector by the parabolic mirror, similar to the technique proposed in [21]. This method provides measuring the angular dependence as it would be observed at infinity. A phase delay was measured by the measurement of interferogram shift (see insertion in Fig. 2) when one of the two mirrors of the interferometer (Fig. 2) was overlapped by the planar structure of metamaterial. Phase delay affected by the substratum was measured separately and then it was subtracted from the total delay.

Geometrically similar structures with cell sizes $a$ = 3, 8.5 and 12 mm (Fig. 1) were prepared for measurements. Fig. 3 presents the dependence between the radiation phase delay $\tau = \frac{2\pi c}{\lambda}(t - t_{substr})$ caused by planar structures as function $\lambda/a$ of the ratio between radiation wavelength $\lambda$ and the size of the structural cell $a$, where $t$ is the time delay by the substratum structure; $t_{substr}$ is radiation time delay by the substratum; $c$ is light velocity.

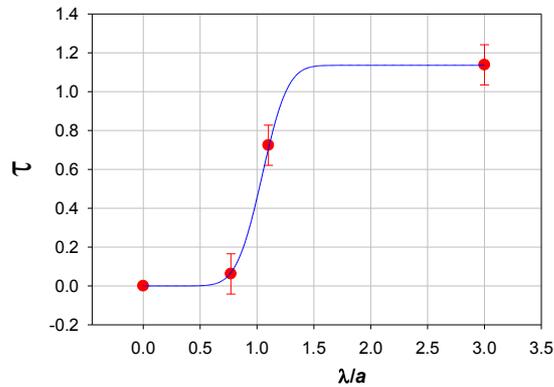

Fig. 3. The radiation phase delay by planar structures as the function of the ratio between radiation wavelength and the size of the structural cell

It is clear from the figure that as it follows from the similarity theory [22], the main effect in the spectrum is observed when the ratio between the radiation wavelength and the size of the structural cell is of the order of unity. In results obtained in [18], the sharp dependence between the refracting angle and the wavelength, right up to the change of the sign of angle, is observed far from this value (unity). In this connection, these results some bewilderment evoke. We believe that this effect is stipulated by the contribution of macroscopic size of the target rather than the influence of the metamaterial structure. Using the results of measurements of the radiation phase delay, the target was prepared in the form of prism the base of which is the right triangle with acute angle of 42° and cell size of 12 mm (Fig. 4). The period of arrangement of two-dimensional structures is 10 mm. This target was used (Fig. 5) to carry out measurements of the radiation refraction characteristics depending on the target angle $\psi$ and observation angles $\theta$. The results obtained are shown on plane $\{\psi, \theta\}$ in which the radiation refraction can be interpreted as refraction with positive refractive index, and regions with negative refractive index (see below). Inside these regions interferograms were measured by the same interferometer as before, which allows to reconstruct the radiation spectra using the inverse Fourier transform by the method described in [23], and a comparison with the initial radiation spectra was made.

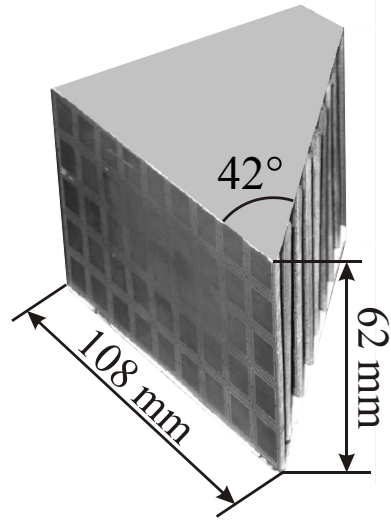

Fig. 4. Configuration of the metamaterial target

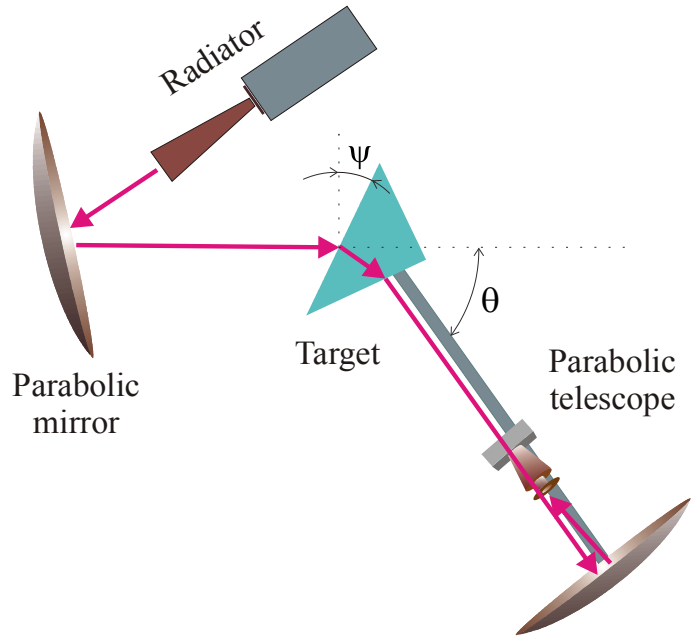

Fig. 5. Experimental scheme for refracted radiation measurement

Below there are three most typical regions which demonstrate both positive and negative refractive index. A clockwise sense corresponds to the positive angular direction. The region with the positive refractive index is shown in geometry N 1 presented in the insertion in Fig. 6(*a*). In this region a weak dependence of the radiation intensity peak on the observation angle and on the target rotation was observed (Fig. 6(*a*)). As a rule, permittivity and permeability and then, the refractive index of metamaterial are not described by scalar function. Nevertheless, one can estimate the scalar refractive index as a rough approximation as it was made in [24] assuming the applicability of Snell law of refraction. For the target type used, the refractive index may be presented in the following expression:

$$n \cdot \sin\left(\varphi - \arcsin\left(\frac{\sin\psi}{n}\right)\right) = \sin(\theta + \varphi - \psi),$$

where $\varphi$ is the angle between the target sides on which refraction occurs.

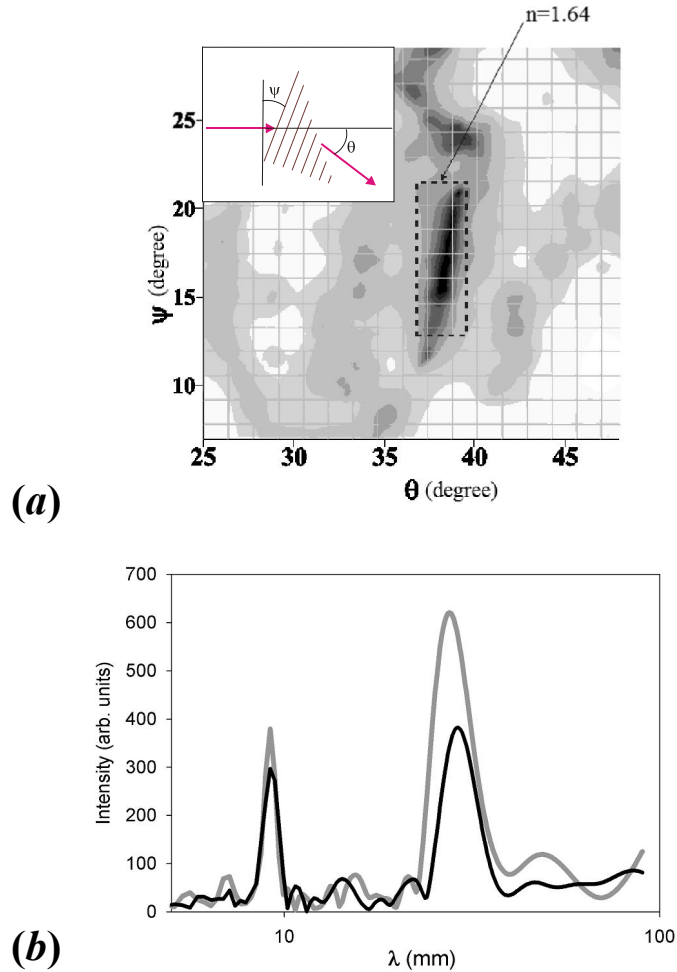

(a)

(b)

Fig. 6. (a) - intensity of refracted radiation on plane $\{\psi,\theta\}$ in geometry N1,
(b) - spectra of the incident radiation (grey line) and refracted radiation (black line)

For geometry N 1 the estimate gives a positive value typical for conventional materials $n = 1.64$. One can see that the refraction region on plane $\{\psi,\theta\}$ is of the local nature. To compare, Fig. 7 shows the angular dependence refracted on the similar Teflon target with refractive index of $n = 1.34$.

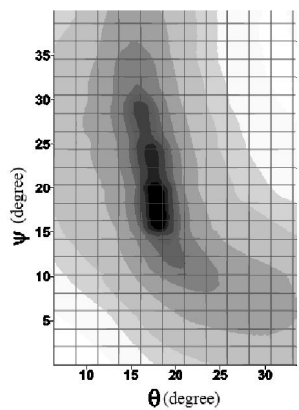

Fig. 7. Intensity of refracted radiation on plane $\{\psi,\theta\}$ for the Teflon target

Spectra of the incident radiation and refracted radiation in metamaterial are given in Fig. 6(b). It is noteworthy that in the geometry N 1 the spectrum of refracted radiation is almost similar to the spectrum of the incident radiation.

Somewhat other situation is observed in geometry N 2 presented in the insertion in Fig. 8(*a*). In this case correlation of the observation angle and the rotation degree of the target is more expressive (Fig. 8(*a*)), although it is not of the direct reflection nature (dashed line in Fig. 8(*a*)).

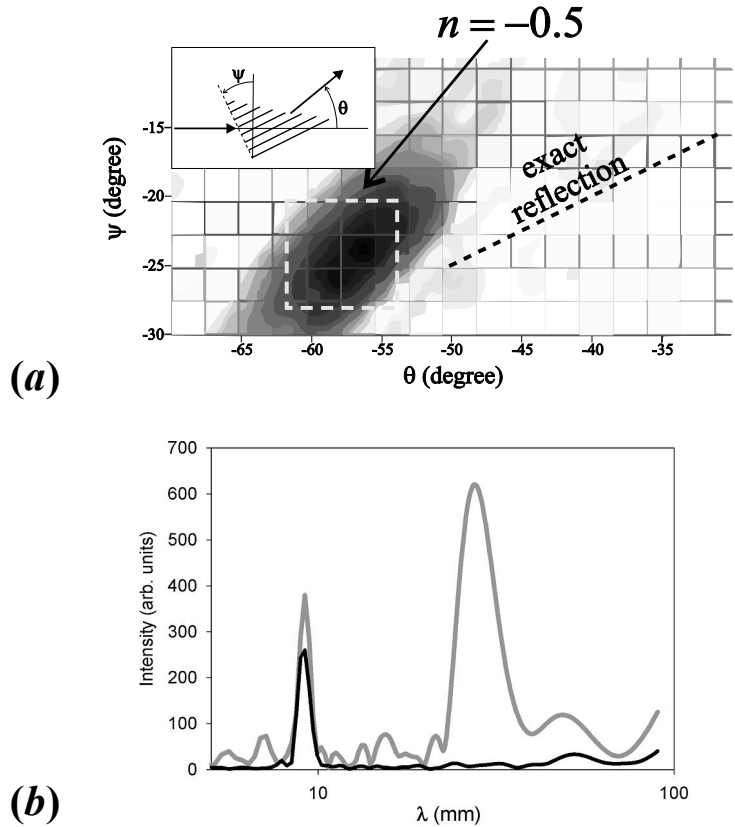

Fig. 8. (*a*) - intensity of refracted radiation on plane $\{\psi, \theta\}$ in geometry N 2,
(*b*) - spectra of the incident radiation (grey line) and refracted radiation (black line)

In this region of the angles the value of the refractive index is already negative $n = -0.5$. Moreover, while comparing the spectrum of the initial radiation and the refracted radiation spectrum (Fig. 8(*b*)) one can see that line $\lambda = 29$ mm is almost completely suppressed, and merely radiation having the wavelength less than the cell size is refracted.

Geometry N 3 shown in the insertion in Fig. 9(*a*) seems to be the most interesting. In this geometry the angular dependence is multimodal (Fig. 9(*a*)); refractive index is negative $n = -1.45$; and the radiation refracts to the backward hemisphere. However, a change of the radiation direction in rotating the target is rather small. It is remarkable that the refracted radiation response fundamentally differs from the previous one. Despite the fact that in the initial radiation the intensity of the long-wave component ($\lambda = 29$ mm) was suppressed, in the refracted radiation it is the main one, while the hard component is completely suppressed (Fig. 9(*b*)). From the point of view of the refracted radiation in conventional materials such radiation behavior is paradoxical.

## 3. Conclusion

The analysis of the results obtained has shown that the radiation behavior in metamaterials is of a complicated nature. It is connected with the fact that optical properties of metamaterials could no longer be considered from the viewpoint of scalar electrodynamic properties of the target material. On the other hand, although electrodynamics with tensor integral parameters of the medium is developed enough [25], there are no exhaustible theory which links the parameters of metamaterial structures and tensor permittivity and permeability, so it is necessary to develop the related region of basic research.

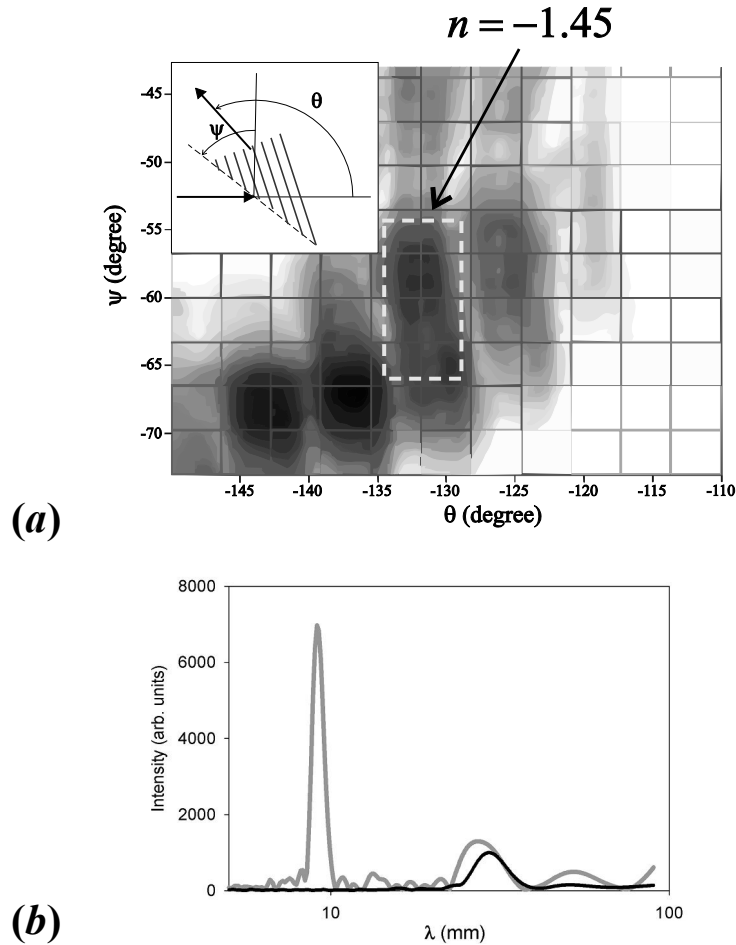

Fig. 9. (*a*) - ntensity of refracted radiation on plane $\{\psi,\theta\}$ in geometry N3,
(*b*) - spectra of the incident radiation (grey line) and refracted radiation (black line)

This work was supported by the Ministry of Education and Science of the Russian Federation within the frame of the Federal target program 'Scientific and Pedagogical Personnel of Innovative Russia' and the Federal target scientific program 0.326.2012.